\documentclass[12pt,letter]{article}

\usepackage{amsmath, amsthm, amssymb}
\usepackage{amssymb,epsfig}
\usepackage{setspace}  

\makeatletter

\gdef\@punct{.\ \ }  
\def\@sect#1#2#3#4#5#6[#7]#8{%
  \ifnum #2>\c@secnumdepth
     \def\@svsec{}
  \else
     \refstepcounter{#1}\edef\@svsec{%
     \ifnum #2>0{{\csname the#1\endcsname}}.\fi%
    \hskip .5em}
  \fi
  \@tempskipa #5\relax
  \ifdim \@tempskipa>\z@
     \begingroup #6\relax
       \@hangfrom{\hskip #3\relax\@svsec}{\interlinepenalty \@M #8\par}
     \endgroup
     \csname #1mark\endcsname{#7}
     \addcontentsline{toc}{#1}{\ifnum #2>\c@secnumdepth\else
          \protect\numberline{\csname the#1\endcsname}\fi#7}
  \else
     \def\@svsechd{#6\hskip #3\@svsec #8\@punct\csname
#1mark\endcsname{#7}
     \addcontentsline{toc}{#1}{\ifnum #2>\c@secnumdepth \else
          \protect\numberline{\csname the#1\endcsname}\fi#7}}
  \fi
  \@xsect{#5}}

\def\@ssect#1#2#3#4#5{\@tempskipa #3\relax
  \ifdim \@tempskipa>\z@
     \begingroup #4\@hangfrom{\hskip #1}{\interlinepenalty \@M
#5\par}\endgroup
  \else \def\@svsechd{#4\hskip #1\relax #5\@punct}\fi
  \@xsect{#3}}

\newcommand{\bx}{\mathbf{x}}
\def\qed{\hskip 3pt \hbox{\vrule width4pt depth2pt height6pt}}

\newtheorem{Lemma}{Lemma}

\newtheorem{Theorem}[Lemma]{Theorem}
\newtheorem{Proposition}[Lemma]{Proposition}

\newtheorem{Definition}[Lemma]{Definition}

\begin{document}

\title{On some upper bounds on the fractional chromatic number of weighted graphs}

\author{Ashwin~Ganesan%
  \thanks{The author is at Amrita University, Coimbatore, India.  Email: \texttt{ashwin.ganesan@gmail.com}. This work was carried out while the author was at the University of Wisconsin at Madison, USA. }
}

\date{}

\maketitle

\vspace{-2.5cm}
\vspace{+2.0cm}

\begin{abstract}
\noindent Given a weighted graph $G_\bx$, where $(x(v): v \in V)$ is a non-negative, real-valued weight assigned to the vertices of G, let $B(G_\bx)$ be an upper bound on the fractional chromatic number of the weighted graph $G_\bx$; so $\chi_f(G_\bx) \le B(G_\bx)$. To investigate the worst-case performance of the upper bound $B$, we study the graph invariant $$\beta(G) ~=~ \sup_{\bx \ne 0}~ \frac{B(G_\bx)}{\chi_f(G_\bx)}.$$

\noindent This invariant is examined for various upper bounds $B$ on the fractional chromatic number.  In some important cases, this graph invariant is shown to be related to the size of the largest star subgraph in the graph.  This problem arises in the area of resource estimation in distributed systems and wireless networks; the results presented here have implications on the design and performance of decentralized communication networks.

\end{abstract}




\setstretch{1.6} 
\section{Introduction}

Let $G=(V,E)$ be a simple, undirected graph on vertex set $V=\{v_1,\ldots,v_n\}$.  Let $\{I_1,\ldots,I_L\}$ be the set of all independent sets of $G$, and let $B = [b_{ij}]$ be the $n \times L$ vertex-independent set incidence matrix of $G$.  Thus, $b_{ij}=1$ if $v_i \in I_j$ and $b_{ij}=0$ if $v_i \not \in I_j$. The chromatic number $\chi(G)$ of $G$ is the value of the program: min $\mathbf{1}^T t$ subject to $Bt \ge 1, t \ge 0, t \in \mathbb{Z}^L$. Equivalently, $\chi(G)$ is the smallest number of independent sets that partition $V$.  Relaxing the condition that $t$ be integral gives the fractional chromatic number \cite{Scheinerman:Ullman:1997} of $G$: $\chi_f(G) = $~min~$\mathbf{1}^T t$ subject to $Bt \ge 1, t \ge 0$.  Now if $G_\bx$ is a weighted graph, where $(x(v):v \in V)$ is a non-negative, real-valued weight assigned to the vertices, the fractional chromatic number $\chi_f(G_\bx)$ of $G_\bx$ is defined as the value of the linear program: min $\mathbf{1}^T t$ subject to $Bt \ge \bx, t \ge 0$.  Equivalently, $\chi_f(G_\bx)$ is the smallest value of $T$ such that each vertex $v$ can be assigned a subset of $[0,T]$ of total length (or measure) $x(v)$, with adjacent vertices being assigned subintervals that are non-overlapping (except possibly at the endpoints of the subintervals).

\bigskip \noindent \textbf{Example.}  Consider the pentagon graph $C_5$ on vertices $v_1,\ldots,v_5$.  Recall that $\chi_f(G)$ is the value of the program: min $\mathbf{1}^T t$ subject to $Bt \ge \bx, t \ge 0$.  Since an independent set in $C_5$ has at most 2 vertices, $\chi_f(C_5) \ge 2.5$.  The assignment $t=(0.5, \ldots, 0.5)$ corresponding to the five maximal independent sets $(\{v_1,v_3\}, \{v_2,v_4\}$, $\{v_3,v_5\}, \{v_4,v_1\}, \{v_5,v_2\})$ is feasible and has optimal value equal to 2.5, yielding $\chi_f(C_5) \le 2.5$.  Thus, $\chi_f(C_5) = 2.5$. In this assignment, subsets of the time interval $[0,2.5]$ are assigned to each vertex such that adjacent vertices are assigned non-overlapping subsets. For example, subsets $[0,0.5]$ and $[1.5,2]$ are assigned to $v_1$, subsets $[0.5,1]$ and $[2,2.5]$ are assigned to $v_2$, subsets $[0,0.5]$ and $[1,1.5]$ are assigned to $v_3$, etc.
\hfill\qed.

\bigskip
The problem of computing the fractional chromatic number of a graph is known to be NP-hard \cite{Grotschel:Lovasz:Schrijver:1981}.  In our work, we study graph invariants associated with the performance of some upper bounds on the fractional chromatic number.  These upper bounds can be efficiently computed; furthermore, they have the property that they can be utilized for resource estimation problems in distributed systems \cite{Ganesan:2009}, \cite{Ganesan:2008}. The results presented here are a combinatorial abstraction of the results on wireless networks in \cite{Ganesan:2009}; also the proofs which are not included there are included here.  The work \cite{Gerke:McDiarmid:2001} discusses a graph invariant associated with the performance of a \emph{lower} bound on the fractional chromatic number.

In the sequel, our notation is standard \cite{Bollobas:2002}.  $\Gamma(v)$ denotes the set of vertices adjacent to G, and $d(v) = |\Gamma(v)|$ is the degree of $v$.  $\Delta(G)$ is the maximum degree of a vertex in $G$.  For $A \subseteq V$, $x(A):=\sum_{v \in A} x(v)$.

\section{Performance of an upper bound based on the greedy coloring algorithm}

One way to color the vertices of $G$ is to pick any ordering of the vertices $v_1,\ldots,v_n$, and to assign to each vertex, in turn, the smallest positive integer not already assigned to its neighbors.  This greedy algorithm produces a coloring of $G$ that uses at most $\Delta+1$ colors, where $\Delta$ is the maximum degree of a vertex in $G$.  As we show next, this bound can be generalized in a straightforward manner to weighted graphs.

Given a weighted graph $G_\bx$, define
$$B_1(G_\bx) := \max_{v \in V} x(v) + x(\Gamma(v)) .$$

\begin{Proposition} For a weighted graph $G_\bx$, we have the upper bound
$$\chi_f(G_\bx) \le B_1(G_\bx). $$

\end{Proposition}

\noindent \emph{Proof}: Let $T:=\max_{v \in V} x(v)+x(\Gamma(v))$.  It suffices to show that it is possible to assign a subset of $[0,T]$ to each vertex such that the length of subintervals assigned to $v$ is at least $x(v)$ and adjacent vertices are assigned non-overlapping subsets. Pick any ordering of the vertices $v_1,\ldots,v_n$. Assign $v_1$ the interval $[0,x(v_1)]$.  Now, assume $v_1,\ldots,v_k$ have already been assigned subsets of $[0,T]$.  Since $x(v_{k+1}) + x(\Gamma(v_{k+1})) \le T$, $x(v_{k+1}) + x (\Gamma(v_{k+1}) \cap \{v_1,\ldots,v_k \}) \le T$.  So it is possible to assign a subset of $[0,T]$ of duration $x(v_{k+1})$ to $v_{k+1}$ which is non-overlapping with the subsets already assigned to its neighbors.  Continuing in this manner with the remaining vertices, we get that $\chi_f(G_\bx) \le T$.
\hfill\qed

\begin{Definition}  The induced star number of a graph $G$ is
defined by
$$ \sigma(G) := \max_{v \in V(G)} \alpha(G[\Gamma(v)]),$$

\end{Definition}
\noindent where $G[V']$ denotes the subgraph of $G$ induced by $V' \subseteq V$ and $\alpha(G)$ denotes the maximum size of an independent set of $G$.  Thus, the induced star number of a graph is the number of leaf
vertices $r$ in the maximum sized star subgraph $K_{1,r}$ of the graph.

\begin{Theorem}\label{thm:row}
$$\sup_{\bx \ne 0} \frac{B_1(G_\bx)}{\chi_f(G_\bx)} = \sigma(G).$$
\end{Theorem}

\noindent \emph{Proof:} Define
$$\beta_1(G) := \sup_{\bx \ne 0} \frac{B_1(G_\bx)}{\chi_f(G_\bx)}.$$
Let $v_1,\ldots,v_{\sigma+1}$ be the vertices of a star subgraph of $G$, where $v_1$ is adjacent to each vertex in the independent set $\{ v_2,\ldots,v_{\sigma+1}\}$.  Consider the weight function $\bx$ that assigns the value $\varepsilon$ to $v_1$, $1-\varepsilon$ to $v_2$ to $v_{\sigma+1}$, and 0 to the remaining vertices.  For this weight $\bx$, $\chi_f(G_\bx)=1$, and $B_1(G_\bx) = \varepsilon+\sigma(1-\varepsilon) = \sigma + \varepsilon(1-\sigma)$, which can be made arbitrarily close to $\sigma$ by making $\varepsilon$ arbitrarily small.  Hence, $\beta_1(G) \ge \sigma(G)$.

To prove that $\beta_1(G) \le \sigma(G)$, pick any weight $\bx$.  Fix any $v \in V$.  Recall that $\chi_f(G_\bx)$ is the value of the program: min $\mathbf{1}^T t$ subject to $Bt \ge \bx, t \ge 0$. An optimal solution to this program gives an assignment of subsets of $[0,\chi_f(G_\bx)]$ to each vertex such that the union of subsets assigned to $\Gamma(v)$ is non-overlapping with the subset assigned to $v$.  Hence, since the maximum size of an independent set in $\Gamma(v)$ is at most $\sigma(G)$, we have that $x(\Gamma(v)) \le \sigma(G)* [ \chi_f(G_\bx) - x(v)]$.  So, $x(v)+x(\Gamma(v)) \le x(v) + \sigma(G)*[ \chi_f(G_\bx) - x(v)] \le \chi_f(G_\bx)~\sigma(G)$.  Hence, $B_1(G_\bx) \le \chi_f(G_\bx) \sigma(G)$.
\hfill\qed

\bigskip \noindent Thus we have shown the fundamental result that for any graph $G$,
$$\sup_{\bx \ne 0}~\frac{\max_{v \in V} x(v) + x(\Gamma(v))}{\chi_f(G_\bx)} = \sigma(G) .$$

\bigskip \noindent In the context of distributed communication networks, this result means that the performance of distributed systems that employ greedy algorithms is limited by the induced star number of the network.  Furthermore, when designing such networks it is desired that the network topology have this quantity to be as close to unity as possible.

\section{Better upper bounds}

In this section we study two more graph invariants associated with the performance of other bounds on the fractional chromatic number.  The final upper bound given below is the strongest of the three bounds.  These upper bounds first appeared in the literature in the context of wireless networks \cite{Hamdaoui:Ramanathan:05}; our new results concern the graph invariants and worst-case performance associated with these bounds.  The proofs are longer and we omit the details due to space limitations; the interested reader can refer to \cite{Ganesan:2008}.

\bigskip \noindent Given a weighted graph $G_\bx$, define
$$B_2(G_\bx) := \max_{v \in V} ~~x(v) (d(v)+1) .$$

\begin{Proposition} For a weighted graph $G_\bx$, we have the upper bound
$$\chi_f(G_\bx) \le B_2(G_\bx). $$

\end{Proposition}

\begin{Lemma}
$$\sup_{\bx \ne 0} \frac{B_2(G_\bx)}{\chi_f(G_\bx)} = \Delta(G)+1.$$
\end{Lemma}

\bigskip \noindent The two bounds $B_1$ and $B_2$ can be combined to get a better upper bound, as follows:
Given a weighted graph $G_\bx$, define
$$B_3(G_\bx) := \max_{v \in V} ~~\min\{~x(v) (d(v)+1)~,~x(v)+x(\Gamma(v))~ \} .$$

\begin{Proposition}
For a weighted graph $G_\bx$, we have the upper bound
$$\chi_f(G_\bx) \le B_3(G_\bx). $$

\end{Proposition}

\begin{Theorem}
$$\frac{1+\sigma(G)}{2} ~\le ~\sup_{\bx \ne 0} \frac{B_3(G_\bx)}{\chi_f(G_\bx)} ~\le ~\sigma(G).$$
Moreover, the lower and upper bounds are tight; the star graphs realize the lower bound, and there exist graph sequences for which the middle quantity approaches the upper bound arbitrarily closely.
\end{Theorem}

Again, we omit the proofs due to space limitations, but the interested reader can refer to \cite{Ganesan:2008}.

Update: The results here were presented at the conference \cite{Ganesan:ICRTGC:2010}. Theorem~\ref{thm:row} above has been published in the journal paper \cite{Ganesan:2010}. A revised version of the technical report \cite{Ganesan:2008} is available online at \cite{Ganesan:2011}.
 {
\bibliographystyle{plain}
\bibliography{winet2}
}
\end{document}